\documentclass[12pt]{article}
\usepackage{amsmath}
\usepackage{graphicx}
\usepackage{natbib}
\usepackage{url} 

\newcommand{\blind}{0}
\usepackage{enumerate}
\usepackage{rotating}
\usepackage{setspace}
\usepackage{graphicx}

\usepackage{enumitem}
\RequirePackage{amsthm,amsmath,amsfonts,amssymb,bm,physics}
\usepackage{subfigure}

\usepackage{setspace}
\usepackage{tabularx} 
\usepackage{booktabs}
\usepackage{authblk}
\usepackage{todonotes}
\usepackage{xcolor}
\usepackage{multirow}
\usepackage{amsmath}
\usepackage{algorithm}
\usepackage{algpseudocode}

\def\mR{\mathbb{R}}
\def\M{\bm{M}}
\def\Q{\bm{Q}}

\newcommand{\bfsym}[1]{\ensuremath{\boldsymbol{#1}}}

\def\bbeta{\bfsym \beta}

             \def\bSigma{\bfsym \Sigma}

\def\btau{\bfsym {\tau}}

\def\mR{\mathbb{R}}
\def\bX{\mathbf{X}}
\def\zero{\mathbf{0}}

\def\boxit#1{\vbox{\hrule\hbox{\vrule\kern6pt\vbox{\kern6pt#1\kern6pt}\kern6pt\vrule}\hrule}}

\usepackage{enumerate}

\newtheorem{theorem}{Theorem}

\newtheorem{prop}{Proposition}
\theoremstyle{remark}
\renewcommand{\bar}{\overline}
\renewcommand{\hat}{\widehat}

\begin{document}

\def\spacingset#1{\renewcommand{\baselinestretch}%
{#1}\small\normalsize} \spacingset{1}


\if0\blind
{
  \title{\bf Bayesian Spatially Clustered Compositional Regression: Linking intersectoral GDP contributions to Gini Coefficients}
  \author{Jingcheng Meng$^1$, Yimeng Ren$^2$,  Xuening Zhu$^2$, Guanyu Hu$^3$\thanks{Jingcheng Meng and Yimeng Ren contributed equally.
    Guanyu Hu is the corresponding author.
    }\hspace{.2cm}\\
    \it\small $^1$ Department of Mathematics and Statistics, Washington University in St Louis. \\
   \it\small $^2$ School of Data Science, Fudan University.\\
    \it\small $^3$ Center for Spatial Temporal Modeling for Applications in Population Sciences, Department of Biostatistics and Data Science,
  The University of Texas Health Science Center at Houston.}
  \maketitle
} \fi

\if1\blind
{
  \bigskip
  \bigskip
  \bigskip
  \begin{center}
    {\LARGE\bf Bayesian Spatially Clustered Compositional Regression: Linking intersectoral GDP contributions to Gini Coefficients}
\end{center}
  \medskip
} \fi

\bigskip
\begin{abstract}
The Gini coefficient is an universally used measurement of income inequality. Intersectoral GDP contributions reveal the economic development of different sectors of the national economy. Linking intersectoral GDP contributions to Gini
coefficients will provide better understandings of how the Gini coefficient is influenced by different industries. In this paper, a compositional regression with spatially clustered coefficients is proposed to explore heterogeneous effects over spatial locations under nonparametric Bayesian framework. Specifically, a Markov random field constraint mixture of finite mixtures prior is designed for Bayesian log contrast regression with compostional covariates, which allows for both spatially contiguous clusters and discontinous clusters. In addition, an efficient Markov chain Monte Carlo algorithm for posterior sampling that enables simultaneous inference on both cluster configurations and cluster-wise parameters is designed. The compelling empirical performance of the proposed method is demonstrated via extensive simulation studies and an application to 51 states of United States from 2019 Bureau of Economic Analysis.
\end{abstract}

\noindent%
\vfill

\newpage
\spacingset{1.75} 
\section{Introduction}
 The Gini coefficient is a widely recognized and commonly employed measure of income inequality \citep{gini1997concentration}.  A high Gini coefficient implies greater inequality in household income. As the disparity in household income has expanded in the United States since the 1980s, policymakers and economists have increasingly focused on identifying the factors that influence income inequality. Understanding the relationships between the Gini coefficient and potential covariates, such as gross domestic product (GDP) \citep{RePEc:spr:sjecst:v:153:y:2017:i:3:d:10.1007_bf03399507}, GDP per capita, the unemployment rate, the single parent household rate \citep{10.1093/ser/mwu001}, and the size of the financial industry \citep{https://doi.org/10.1111/meca.12165}, will contribute to a more comprehensive understanding of the factors affecting the Gini coefficient and, consequently, income inequality. Furthermore, it has been suggested that industry composition is a potential factor that influences income inequality \citep{doi:10.1146/annurev.soc.33.040406.131755}. \citet{10.2307/41219998} used a multiple regression model to investigate the relationship between the Gini coefficient and the composition of the industry. Their findings indicate that a decrease in the size of the construction sector and an increase in the size of the FIRE (Finance, Insurance, and Real Estate) sector lead to a higher Gini coefficient and increased income inequality. To elaborate, industry composition is often represented by the contributions of different sectors to GDP, and can be considered compositional data when derived from various components. 
 Compositional data modeling has been widely discussed in the context of log-contrast regression models under various frameworks \citep{10.2307/43957087,10.2307/26452944}, adhering to the principle of subcompositional coherence in regression analysis. To model intersectoral GDP contributions across the spatial domain, it is essential to combine compositional data modeling with spatial analysis. However, research on log-contrast regression in the spatial domain remains limited.

 The heterogeneity pattern in spatial data in different locations is commonly observed in practice \citep{gelfand2003spatial,lee2017cluster,leespatial2019,li2019spatial,ma2019bayesian,hu2020bayesian,geng2020bayesian}. Existing approaches that account for such patterns in regression models can be put into two
major categories. The first one is to incorporate spatial random effects in regression models with constant coefficients via parametric ways \citep{datta2019spatial} or nonparametric ways \citep{gelfand2005bayesian}.  
Another important approach, instead of assuming all covariate effects are constant, is spatially varying
coefficient model (SVCM), 
 which allows the coefficients of the covariates to change with the locations have been widely used. 
The SVCMs are flexible tools for studying spatial regression problems by incorporating spatial heterogeneity and nonstationarity. 
 However, they may face the risk of over-fitting the data and low estimating efficiency, as each spatial location is associated with a distinct coefficient vector \citep{zhang2022learning}. 
The number of unknown parameters in the model would increase dramatically as the location size grows, and thus it can easily lead to over-parametrization. This issue poses a great challenge in the context of large spatial data. The conventional SVCMs suffer from the following limitations.
Specifically, the heterogeneity pattern in such type of model fails to consider similar regression patterns inside neighborhoods and assign distinct coefficients while ignoring the spatial connections \citep{jiangclustering}.
A thorough understanding of subgroups or clusters is crucial for helping practitioners develop local policies and economic development strategies. As a result, the development of spatial clustering methods has emerged as a central research topic in the social and regional economic fields. Consequently, proposing a spatially clustered regression model for compositional predictors, which links intersectoral GDP contributions to Gini coefficients, addresses intermediate needs in social science and economics.

In summary, there are three challenges for spatially clustered regression for compositional predictors. Firstly, it is necessary to consider location information in clustering process. For instance, Missouri and Kansas, which have very similar industrial structures and predictor effects, are highly likely to belong to the same group. 
However, most existing clustering methods, such as $k$-means and mixture regressions, do not incorporate spatial information. 
 Second, both locally contiguous clusters and globally discontiguous clusters should be considered. For example, while Texas and California are geographically distant, they share similar demographic information, such as population and income, and may belong to the same cluster. In other words, spatial contiguous constraints cannot dominate the global cluster configuration. Lastly, determining the number of clusters is an important consideration for regression models with clustered coefficients. Many existing methods use information criteria to decide the number of clusters and then estimate the cluster configurations \citep{heaton2015nonstationary}. Such a two-step procedure may ignore the uncertainty of estimating the number of clusters in the first stage and is prone to increase incorrect cluster assignments in the second stage.

The primary objective of this paper is to address these challenges by introducing a novel Bayesian compositional regression model with clustered coefficients to learn the relationship between Gini coefficients and intersectoral GDP contributions between different states in the US. We first introduce a Helmert transformation \citep{watson2006computing} for regression coefficients in log-contrast regression. The Helmert transformation offers two key benefits: it automatically omits the redundant dimension in regression coefficients and enables the convenient extension of nonparametric Bayesian methods for clustering coefficients of compositional predictors. Nonparametric Bayesian methods have been widely applied in various fields due to their intuitive probabilistic interpretation and elegant computational solutions, such as the collapsed Gibbs sampler \citep{neal2000markov}. However, nonparametric Bayesian methods for spatial log-contrast regression have received limited attention to date.

The contributions of this paper are in three-folds. First, the proposed Bayesian nonparametric compositional regression method is able to leverage spatial information without pre-specifying the number of clusters. The proposed method guarantees the local contiguous constraints and global discontinuous clusters at the same time. In fact, this idea and our proposed approach are widely applicable to general compositional data analysis such as biology and environmental science, and provide a valuable alternative to the existing literature that mainly relies on penalized approach \citep{ma2017concave,li2019spatial,su2016identifying,qian2016shrinkage,su2018identifying} or finite mixture model \citep{huang2012mixture}. Second, by using a Bayesian framework, the probabilistic interpretations of clustering results are easily obtained. Thirdly, posterior inference of the 
cluster-wise parameters and clustering information (both the number of clusters and clustering configurations) will be efficiently and conveniently implemented on blessings of the developed posterior sampling scheme without complicated reversible jump MCMC or allocation samplers.

The rest of this paper is organized as follows. In Section \ref{sec:data}, we will briefly give an overview of our motivating data application. In Section \ref{sec:method}, we will review the log-contrast regression and propose a Bayesian log-contrast regression model with clustered coefficients under Markov random field constraint mixtures of finite mixtures prior. 
In Section \ref{sec:bayes_inference}, we will derive the Bayesian inference procedures for our proposed methods, including the MCMC algorithm, post-MCMC estimation, and the model selection criterion for tuning parameters. Extensive simulation studies are presented in Section~\ref{sec:simu} to investigate the empirical performance of our approach. We apply our method to analyze the state-wise economics data from the Bureau of Economic Analysis in Section~\ref{sec:app} and conclude with a discussion of future directions in Section~\ref{sec:discussion}.


\section{Motivating Data}\label{sec:data}
Our motivating data come from the 2019 Bureau of Economic Analysis, U.S. Department of
Commence, which includes Gini coefficients, intersectoral GDP contributions data, household income per capita, and unemployment rate.  All data are recorded for the 50 states plus Washington, DC. We will refer to them with ``51 states" for simplicity in the rest of this paper.

\begin{figure} [H]
\centering
	\subfigure[]{
	\includegraphics[scale=0.4]{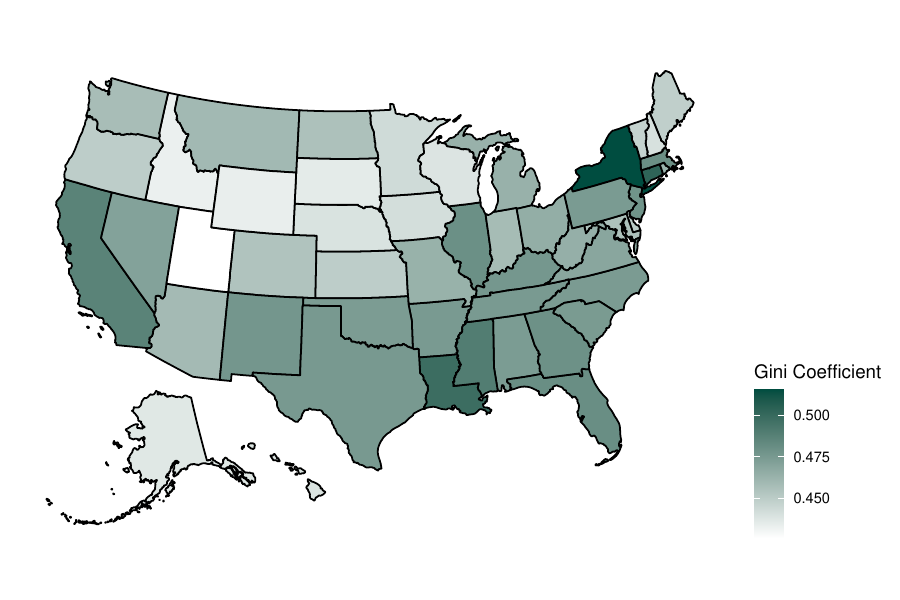}
	}
	\subfigure[]{
	\includegraphics[scale=0.4]{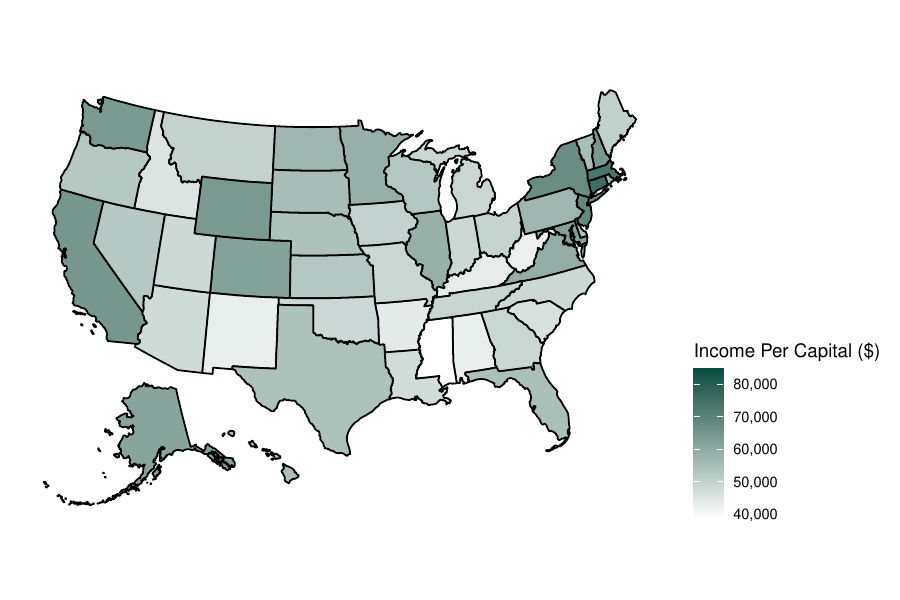}
	}
 \subfigure[]{
	\includegraphics[scale=0.4]{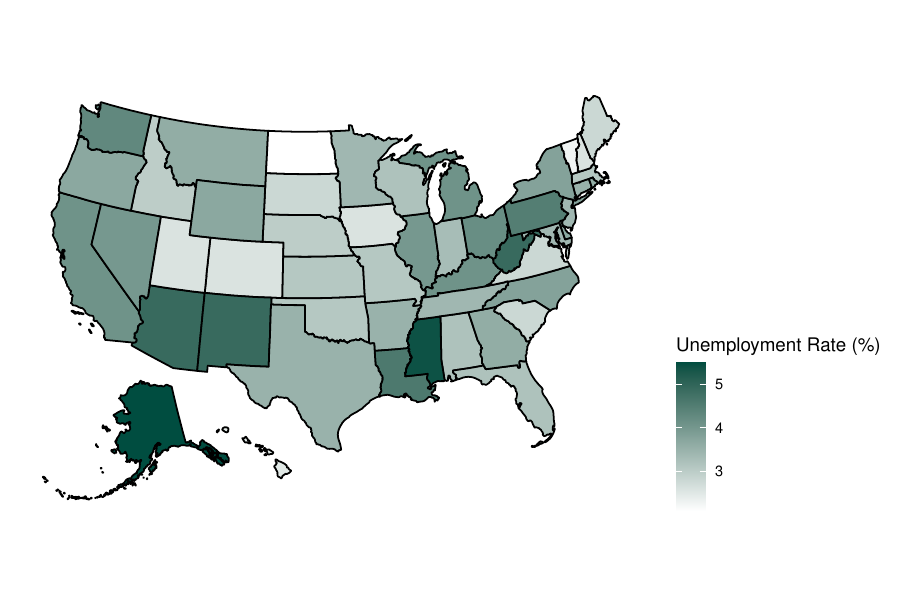}
	}	\caption{Descriptive statistics on the US map: (a) Gini Coefficient; (b) Household Income Per Capital; (c) Unemployment Rate. } \label{fig:Motivating Data}
\end{figure}

In the following section, we present these descriptive statistics visually. 
Figure \ref{fig:Motivating Data}(a) illustrates the Gini coefficient. Utah exhibits the lowest income inequality with a Gini coefficient of 0.427, while Washington DC displays the highest income inequality with a Gini coefficient of 0.512. Notably, regions along the east coast, west coast, and southern United States exhibit greater income inequality compared to other areas. Household income per capita is illustrated in Figure \ref{fig:Motivating Data}(b) portrays household income per capita. Mississippi records the lowest average household income, whereas Washington DC boasts the highest. Regions such as the west coast and the New England area demonstrate higher average household incomes compared to other parts of the country. Figure \ref{fig:Motivating Data}(c) provides insights into the unemployment rate across the United States. North Dakota reports the lowest unemployment rate at 2.1, while Alaska and Mississippi register the highest rates, successively. 

\begin{figure} [h]
	\centering
	\includegraphics[scale=0.36]{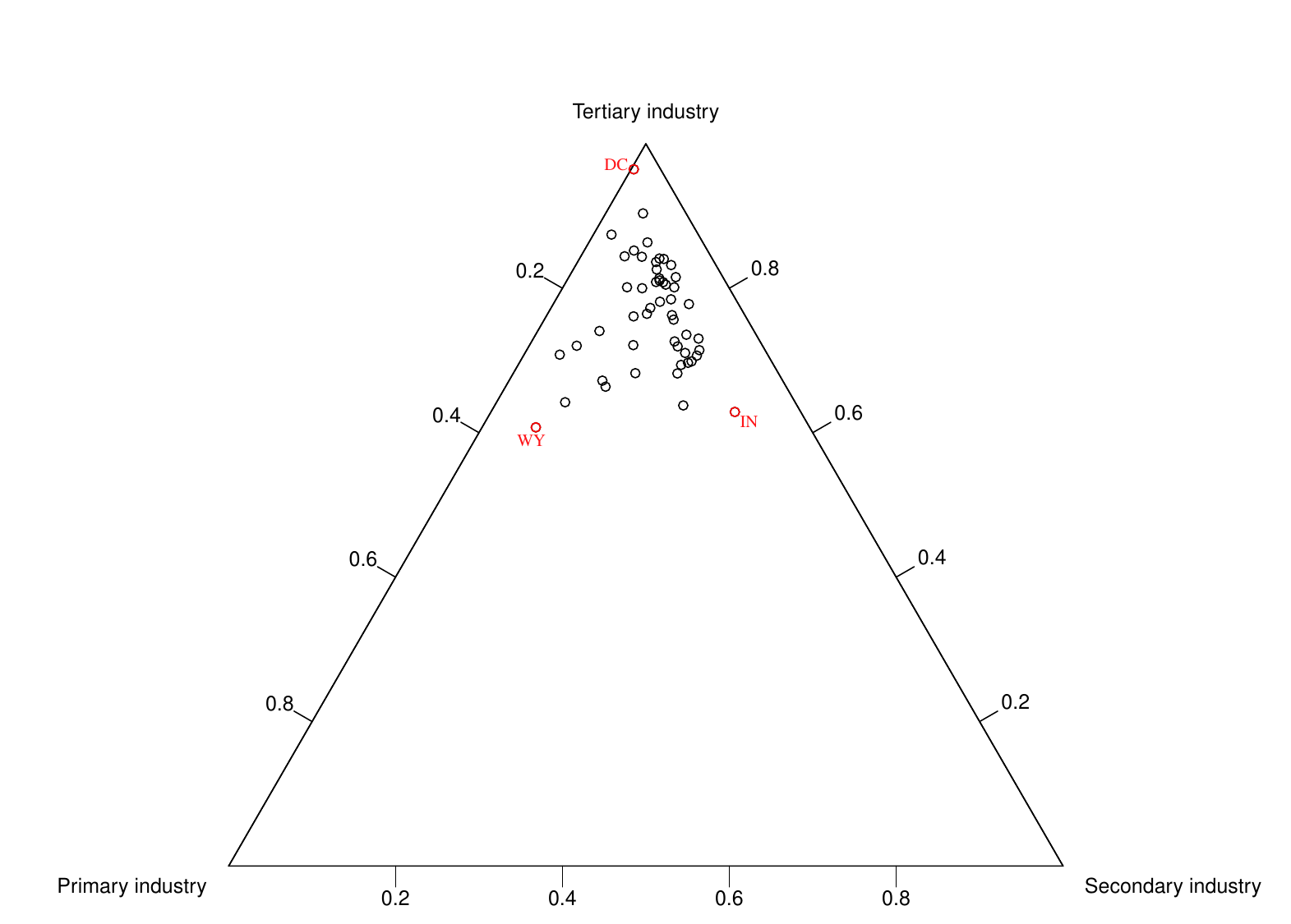}
	\caption{Ternary diagram of the U.S. 2019 intersectoral GDP contributions data for 51 states.} \label{fig:ternary}
\end{figure}

Intersectoral GDP contributions data shows the proportion of each industry's contribution to the GDP of the US. The industry list of the intersectoral GDP contributions data are given in Table~\ref{tab:industry_list}.
To visualize the proportion of different industries for each state, we 
employ a ternary diagram, where we categorize the above industries into primary, secondary and tertiary industries and calculate the corresponding proportion in Figure \ref{fig:ternary}.
In Figure \ref{fig:ternary}, it is shown that in general, the contribution of the tertiary industry in all states is the highest, but the compositions vary between different states. For example, for Wyoming (WY), the contribution of the primary industry is 33\%, which is the highest among states. Similarly, for Indiana (IN) and the District of Columbia (DC), the contributions of the secondary and tertiary industries are the highest, respectively. Homogeneity patterns can be seen from the diagram as well. For some states, including DC, the contribution of the tertiary industry is extremely high, while the contribution of the primary industry is extremely low. However, for some states in the middle, the primary and secondary industries contribute roughly equally to the GDP.



\begin{table}[H]
\centering
\caption{Mean and median values of GDP contribution for each industry across all 51 states in the United States}\label{tab:industry_list}
\footnotesize
{
\begin{tabular}{lcc}
    \toprule Industry List&Mean&Median\\
    \midrule
   1) Agriculture, forestry, fishing
and hunting&0.0129&
0.0075\\

2) Mining, quarrying, and oil and gas extraction&
0.0222&
0.0032
\\

 3) Utilities&
0.0177&
0.0169
\\

4) Construction&
0.0435&
0.0410
\\

5) Manufacturing&
0.1105&
0.1045
\\

6) Wholesale trade&
0.0568&
0.0559
\\

7) Retail trade&
0.0583&
0.0567
\\

8) Transportation and warehousing&
0.0374&
0.0332
\\

9) Information&
0.0371&
0.0302
\\

10) Finance&
0.2041&
0.2041
\\

11) Professional and business services &
0.1152&
0.1144
\\

12) Educational services,
health care, and social assistance&
0.0929&
0.0918
\\

 13) Arts, entertainment, recreation, accommodation, and food services &
0.0453&
0.0392
\\

14) Other services except government and government enterprise&
0.0229&
0.0223
\\

 15) Federal civilian&
0.0304&
0.0194
\\

16) State and local
spending&
0.0921&
0.0894\\
\bottomrule

\end{tabular}
}
\end{table}

The intersectoral GDP contribution data are compositional data. Compositional data indicates the the relative amount of proportions of a whole and the data sums up to a constant value. The sample space of compositional data, called a simplex, is denoted by $\mathcal{S}^{k-1}=\{\bm{x}=[x_1,...,x_k]\in \mathbb{R}^k|x_i\geq 0,i=1,...,k;\sum_{i=1}^k x_i=1\}$ where $k$ is the number of components. The intersectoral GDP shows the proportion of each industry's contribution, and the contributions data of each state sum up to 1. If we do regression analysis directly with compositional data as a predictor, the compositional data will lead to the parameter identifiable issue in the linear regression. For example, if the real model is $y=1+2x_1+x_2$ where $y\in R$ and $\{x_1,x_2\}\in \mathcal{S}^1$. We may obtain $y=2+x_1$ or $y=3-x_2$ as our estimated model which is consistent with the data, but the coefficients of $x_1$ and $x_2$ are not in agreement with our real model. In this paper, we will follow  \cite{aitchison1982statistical} and introduce log-contrast regression model, using Hessian matrix to avoid the parameter identifiable issue, which will be presented in the next section.

\section{Methodology}\label{sec:method}


\subsection{Bayesian Log-Contrast Regression for Compositional Covariates}
 In this section, we will first describe a log-contrast regression model for compositional predictors. To identify potential links between intersectoral GDP contributions and Gini coefficients, a log-contrast regression with compositional predictors will be discussed. Suppose we observe $n$ independent observations of a {continuous type} response
$y_i \in \mathbb{R}$, a $k$ dimensional compositional predictor $\tilde{\bm{X}}_{1i} = [x^{(1)}_{i1},\ldots,x^{(1)}_{ik}]^\top$, such that $\tilde{\bm{X}}_{1i}\in \mathcal{S}^{k-1}$, and another $p$ dimensional non-compositional predictor $\bm{X}_{2i}=[x^{(2)}_{i1},...,x^{(2)}_{ip}]^\top \in \mathbb{R}^p$.  Denote $\bm{y}=[y_1,\cdots,y_n]^\top $, $\tilde{\bm{X}}_1=[\tilde{\bm{X}}_{11},...,\tilde{\bm{X}}_{1n}]^\top $ and $\bm{X_2}=[\bm{X}_{21},...,\bm{X}_{2n}]^\top $. A linear model for $\bm{y}$ could be expressed as
\begin{equation*}
    \bm{y}=\tilde{\bm{X}}_{1}\bar{\bm{\beta}}+\bm{X}_{2}\bm{\eta}+\bm{\epsilon}    
\end{equation*}
where $\bar{\bm{\beta}}$ is the $k$ dimensional regression coefficient, $\bm{\eta}$ is the $p$ dimensional regression coefficient, $\bm{\epsilon}\in \mathbb{R}^n$ is an $n$ dimensional random error vector with zero mean and variance $\sigma^2$.
Ignoring the simplex structure of $\tilde{\bm{X}}_{1}$ would cause the issue of parameter identification in the
linear regression of $\bm{y}$ on $\tilde{\bm{X}}_1$. One naive ``remedy'' is to exclude an arbitrary component of
the compositional vector in the regression, but this may lead to a method that is not
invariant to the choice of the removed component, since it affects both prediction and
selection. Consequently, this would pose difficulty in properly interpreting and inferring the model. Following \cite{lin2014variable}, we use a log-ratio transformation of the compositional data,
such that the transformed data admit the familiar Euclidean geometry in $\mathcal{R}^{K-1}$. Specifically,
\begin{equation}
    \bm{y}=\bm{Z}\tilde{\bm{\beta}}+\bm{X}_{2}\bm{\eta}+\bm{\epsilon}, \qquad  \text{s.t.}\quad \sum_{j=1}^K\tilde{\beta}_{j}=0,
\end{equation}
where $\tilde{\bm{\beta}}=[\tilde{\beta}_{1},...,\tilde{\beta}_{k}]$
is the regression coefficient vector for the transformed design matrix $\bm{Z}$, which is the log-ratio transformation of $\tilde{\bm{X}}_{1}$. 
 In order to remove the constraint of $\tilde{\bbeta}$, a Helmert transformations is considered as 
\begin{equation}
   \bm{\beta}=\bm{H}\tilde{\bm{\beta}},
   \label{eq:finalmodel}
\end{equation}
where $\bm{H}\in \mR^{(K-1)\times K}$ is the Helmert sub-matrix \citep{lancaster1965helmert} with the the first row omitted. 
The Helmert transformation matrix $\bm{H}$ is not a full row rank matrix.  We propose an orthogonal projection \citep{maynard2005drawing} by following the theorem, which provides a way to transform $\bbeta$ to $\tilde \bbeta$.

\begin{theorem}
\label{the:orth_prior}
Let $\bm{H} = \bm{F}\bm{Q}$ be a full rank decomposition
of $\bm{H}$. Then $\bm{F}\in \mR^{(K-1)\times r}$ is a full column rank matrix and $\bm{Q}\in \mR^{r\times K}$ is a full row rank matrix. 
 Write $\M = (\Q^\top,(\Q^\bot)^\top)^\top\in\mR^{K\times K}$, where $\bm{Q}^{\bot}\in\mR^{(K-r)\times K}$ is the orthogonal compliment of $\Q$ satisfying $\Q(\Q^{\bot})^\top = \zero$.
Let $\tilde\M = (\tilde{\bm{M}}_1,\tilde{\bm{M}}_2)=\bm{M}^{-1} $ with $\tilde\M_1\in \mR^{K\times (K-1)}$.
Then
the inverse transformation of $\bm{\tilde{\beta}}$ is 
\begin{equation}
	\tilde{\bm{\beta}}=\tilde{\bm{M}}_1\bm{\beta}.
\label{eq:fuse_prior}
\end{equation}
 \end{theorem} 
 Based on the Theorem \ref{the:orth_prior},  we have the following linear regression model instead of a regression model with linear constraint
\begin{equation}
\bm{y}=\bm{X}_{1}\bm{\beta}+\bm{X}_{2}\bm{\eta}+\bm{\epsilon},
\label{eq:new regression}
\end{equation}
where $\bm{X}_1=\bm{Z}\tilde{\bm{M}}_1$. Thus, we eliminate the problem of dealing with compositional covariates by log-ratio transformation and omit the redundant dimension of $\bbeta$ using the Hermert matrix. For \eqref{eq:new regression}, a joint prior for $\bm{\beta}$, $\bm{\eta}$ and $\sigma^2$ can be used to complete the Bayesian model.


\subsection{Bayesian Spatially Clustered Regression}

For many spatial economics data, regions may share the same predictor effects with their
nearby regions. In the meanwhile, regions may share similar parameters regardless of their geographical distances, due
to the similarities of regions' demographical information such as population and tax rate. A spatially varying pattern for covariate effects may
not be always valid. Based on the homogeneity pattern, we focus on the clustering of regression coefficients of compositional predictors. In our setting, we assume that the $n$ regression coefficient vectors can be clustered into $k$ groups. For known $k$ settings, a finite mixture model is a natural solution for probabilistic clustering. However, the performance of the estimation of cluster
assignments highly relies on the pre-specified number of clusters, it may ignore the uncertainty in the number of clusters and cause redundant cluster assignments. The Bayesian nonparametric method is a natural solution for simultaneously estimating the number of clusters and cluster configurations. Suppose each individual $i$ carries a latent group $z_i$. The conditional distributions of $z_1,\ldots,z_n$ could be formulated as Chinese restaurant process \citep[CRP,][]{pitman1995exchangeable}.

However, the CRP has been shown to produce redundant tail clusters, causing inconsistency in estimation for the number of clusters even with a large sample size. In addition, the spatial information among the states is not taken into consideration for the CRP. 
Another modification of the Dirichlet process mixture model is proposed, known as the mixture of finite mixtures (MFM) model, to mitigate the inconsistency issue \citep{miller2018mixture}. 
The MFM model can be formulated as
$$k\sim p(\cdot),\quad (\pi_1,\ldots,\pi_k)|k \sim \text{Dirichlet}(\gamma,\ldots,\gamma),\quad z_{i}|k,\bm{\pi}\sim \text{Cat}(k,\bm{\pi}),\quad i=1,\ldots,n,$$
with $p(\cdot)$ being the Poisson distribution function truncated to be positive (i.e., $k-1 \sim \text{Poisson}(\lambda)$), and $\text{Cat}(k,\bm{\pi})$ is k-dimensional categorical distribution.

The MFM model can also be formulated as a similar restaurant process:
\begin{eqnarray}
P\left(z_i=c|\bm{z}_{-i}\right)\propto 
\begin{cases}
      |c|+\gamma,&\text{ at an existing  label $c$}\\
      V_n(K^{*}+1)/V_n(K^{*})\gamma, &\text{ if $c$ is a new label}
\end{cases},
\label{eq:MFM}
\end{eqnarray} 
where $K^{*}$ is the number of existing clusters and the coefficient $V_n(w)$ is computed as
\begin{align*}
    V_n(w)=\sum_{k=1}^{\infty}\frac{k(w)}{(\gamma k)^{(n)}}p(k),
\end{align*}
where  $k^{
(m) }= k(k + 1)\ldots(k + m-1)$
and $k(m) = k(k-1)\ldots(k- m+1)$, with $k^{
(0)} = 1$ and $k(0) = 1$. The coefficient $V_n(w)$ also slows down the rate of introducing new clusters to existing ones, which can avoid having many tiny extraneous groups in the cluster. Another important consideration for spatial homogeneity learning is borrowing spatial information.  Our remedy is combining Markov random field \citep[MRF,][]{orbanz2008nonparametric} with MFM. The dependence structure of different variables can be represented by a graph, with vertices representing random variables and an edge connecting two vertices indicating statistical dependence. The Markov random field constrained MFM (MRFC-MFM) consists of an interaction term modeled by an MRF cost function that captures spatial interactions among vertices and a vertex-wise term modeled by an MFM. 
Denote $\bm{\theta}(s_i)$ by the parameter of the location $i$.
The joint prior for $\{\theta(s_i): 1\le i\le n\}$ for MRFC-MFM is given as 
\begin{equation}
    \pi(\bm{\theta}(s_1),\ldots,\bm{\theta}(s_n))\propto P(\bm{\theta}(s_1),\ldots,\bm{\theta}(s_n))M(\bm{\theta}(s_1),\ldots,\bm{\theta}(s_n)|\mathcal{G}),
\end{equation}
with $P(\bm{\theta}(s_1),\ldots,\bm{\theta}(s_n))$ as the part of the joint prior distribution for $\bm{\theta}(s_1),\ldots,\bm{\theta}(s_n)$ induced by MFM, and $M(\bm{\theta}(s_1),\ldots,\bm{\theta}(s_n)|\mathcal{G})$ as anther part of joint prior distribution which is induced by Markov random field given a graph structure $\mathcal{G}$.  The pre-specified $\mathcal{G}$ is defined as an unweighed graph with
vertices $V_\mathcal{G} = (v_1,\ldots,v_n)$ representing random variables at~$n$
spatial locations,  $E_\mathcal{G}$ denoting a set of edges representing
statistical dependence among vertices.
By the Hammersley-Clifford theorem \citep{clifford1971markov}, the corresponding conditional distributions of $M(\bm{\theta}(s_1),\ldots,\bm{\theta}(s_n)|\mathcal{G})$ enjoy the Markov property, i.e., $M(\bm{\theta}(s_i)|\bm{\theta}(s_{-i}))=M(\bm{\theta}(s_i)|\bm{\theta}(s_{\partial (i)}))$, where $\bm{\theta}(s_{-i})=(\bm{\theta}(s_{1}),\ldots,\bm{\theta}(s_{i-1}),\bm{\theta}(s_{i+1}),\ldots,\bm{\theta}(s_n))$ and $\partial (i)$ denotes the set of neighborhood locations of $s_i$ given the graph $\mathcal{G}$.

\begin{prop}
\label{thm:MRF-MFM} Let $n_k^{(-i)}$ denote the size of the $k$-th
cluster excluding $\bm{\theta}(s_i)$, $K^*$ denote the number of clusters excluding the
$i$-th observation, $\bm{\theta}^*_1,\ldots, \bm{\theta}^*_{K^*}$ denote $K^*$ distinguished parameters and assume $M(\bm{\theta}(s_1),\ldots,\bm{\theta}(s_n)|\mathcal{G})=
\frac{1}{Z_H}\text{exp}\{-H(\bm{\theta}(s_1),\ldots,\bm{\theta}(s_n)|\mathcal{G})\}$, where $\frac{1}{Z_H}$ is the normalizing constant. The conditional distribution of an MRFC-MFM takes the
form
\begin{equation}
\Pi(\bm{\theta}(s_i)\mid\bm{\theta}(s_{-i})) \propto \sum_{k=1}^{K^*}
(n_k^{(-i)}+\gamma)\frac{1}{Z_H}\exp(-H(\bm{\theta}(s_i)\mid
\bm{\theta}(s_{-i})))\delta_{\bm{\theta}^*_k}
(\bm{\theta}(s_i)) + \dfrac{V_n(K^*+1)}{V_n(K^*)}\dfrac{\gamma}
{Z_H}G_0(\bm{\theta}(s_i))
\label{eq:cond_dist}
\end{equation}
 where 
\begin{equation}
   H(\bm{\theta}(s_i)\mid
\bm{\theta}(s_{-i})) =  -\lambda\sum_{\ell\in \partial (i)}I(z_{\ell}=z_i).
    \label{eq:cost}
\end{equation}
$\delta_{\bm{\theta}^*_k}
(\bm{\theta}(s_i))$ is the distribution concentrated at a single point, $\bm{\theta}^*_k$, and a base measure $G_0$ is defined the same as the Dirichlet process \citep{neal2000markov}. 
\end{prop}

In Proposition \ref{thm:MRF-MFM}, $\lambda$ is a spatial smoothness parameter.
A larger value of $\lambda$ indicates stronger spatial smoothing. Proposition \eqref{thm:MRF-MFM} gives a similar condition distribution with traditional Chinese resturant processs.
Combining \eqref{eq:cond_dist} and \eqref{eq:cost}, we can have a similar conditional distribution for $z_1,\ldots,z_n$ as
\begin{eqnarray}
P\left(z_i|\bm{z}_{-i}\right)\propto 
\begin{cases}
      (|c|+\gamma)\text{exp}[\lambda\sum_{l\in \partial_s (i)}I(z_l=z_i)], &\text{at an existing labeled c}\\
      V_n(K^{*}+1)/V_n(K^{*})\gamma, &\text{if c is new cluster}
\end{cases}.
\end{eqnarray}
The above urn scheme offers a similar Chinese restaurant process interpretation \citep{neal2000markov} of the proposed prior: the probability of customer~$i$ sitting at a table
depends not only on the number of existing customers seated at that table
but also on spatial relationships between the $i$-th
customer with existing customers. Compared with traditional MFM and CRP, this
P\'{o}lya urn scheme will let nearby states have a higher
probability being clustered together when $\lambda \neq 0$.
This will enforce the locally contiguous clusters.
The globally discontiguous clusters will be learned from the data itself. Compared with existing Bayesian approaches \citep{lu2007bayesian,li2015bayesian,gao2023spatial,aiello2023detecting} for spatial clustering detection, the clustering labels can be inferred directly by the posterior estimates without any FDR-based post selection procedures.

\subsection{Bayesian Hierarchical Model}
Consider the following specifications of the data models for linking state-specific Gini coefficients ($y_i, i=1,\ldots,n$) and both transformed compositional predictors $\bm{X}_{1i}=\bm{Z}_i\tilde{\bm{M}}_1$ and non-compositional predictors $\bm{X}_{i2}$ 
\begin{equation}
\label{eq:data_model}
    y_i|\bm{X}_{1i},\bm{X}_{2i},\bm{\beta}(s_i),\bm{\eta} \sim \mathcal{N}(\bm{X}_{1i}\bm{\beta}(s_i)+\bm{X}_{2i}\bm{\eta},\sigma^2(s_i)),
\end{equation}
where $\bm{\beta}(s_i)$ is spatially varying coefficients for compositional predictors, $\bm{\eta}$ is spatially constant coefficients for non-compositional predictors, and $\sigma^2(s_i)$ is spatially varying variance.

For spatially constant coefficients $\bm{\eta}$, a multivariate normal prior is given as 
\begin{equation}
\label{eq:prioreta}
    \bm{\eta}\sim \mathcal{N}_p(\bm{\eta}_0,\bm{V}_0),
\end{equation}
where hyperparameters $\bm{\eta}_0=\bm{0}_p$ and $\bm{V}_0=100\bm{I}_p$. In order to capture the spatially clustered pattern of regression coefficients for compositional predictors, a MRFC-MFM prior is proposed for $\bm{\theta}(s_i)=(\bm{\beta}^\top(s_i),\sigma^2(s_i))^\top$ with Normal-Inverse-Gamma (NIG) base distribution as 
\begin{equation}
\label{eq:jointnigprior}
    \begin{split}
    \bm{\theta}(s_1),\ldots, \bm{\theta}(s_n)&\sim M(\bm{\theta}(s_1),\ldots, \bm{\theta}(s_n)|\mathcal{G})\prod_{i=1}^n G(\bm{\theta}(s_i)),\\
    G(\bm{\theta}(s_i))| \pi_1,\ldots,\pi_K, K &\sim \sum_{\ell=1}^K \pi_\ell \text{NIG}(\bm{\tau}_0,\Sigma_0,a_0,b_0),\\
    \pi_1,\ldots,\pi_K|K &\sim \text{Dirichlet}(\gamma,\ldots,\gamma),\\
    K-1&\sim \text{Poisson}(\zeta),
    \end{split}
\end{equation}
where $\bm{\tau}_0=\bm{0}, \bm{\Sigma}_0=\bm{I}, a=0.01, b=0.01$, $\bm{\eta}_0=\bm{0} $ and $\bm{V}_0=100\bm{I_p}$ are the hyperparameters for NIG distribution, and $\zeta$ is pre-specified parameters for Poisson distribution. $M(\bm{\theta}(s_1),\ldots, \bm{\theta}(s_n))$ is part of joint prior induced by Markov random field given a graph structure $\mathcal{G}$. The $\mathcal{G}$ is the spatial adjacency structure in our simulation and application. In the rest of paper, we choose $\zeta=1$ and $\gamma=1$ as \citep{miller2018mixture}. Combining \eqref{eq:data_model}, \eqref{eq:prioreta}, and \eqref{eq:jointnigprior}, we finish our hierarchical model.

\section{Bayesian Inference}\label{sec:bayes_inference}

In this section, we will introduce the MCMC sampling
algorithm, post-MCMC inference method, and Bayesian
model selection criterion.

\subsection{MCMC Algorithm}

Our goal is to sample from the posterior distribution
of the unknown parameters $K$, $\boldsymbol{z}=(z_1,...,z_K), \boldsymbol{\beta}=({\boldsymbol{\beta}_1,...,\boldsymbol{\beta}_K}),\boldsymbol{\sigma^2}=({{\sigma_1^2},...,{\sigma_K^2}}), \boldsymbol{\eta}=(\eta_1,...,\eta_p)$. 
The marginalization
over $K$ can avoid complicated reversible jump MCMC algorithms
or even allocation samplers. For posterior computation we use a Gibbs sampler defined by the following propositions.

\begin{prop}
\label{prop1}
The full conditional distributions of $z_i$'s are given
\begin{align*}
P(z_i|\bm{z}_{-i},\bm{\beta},\bm{\sigma}^2,\bm{\eta}) \propto\left\{\begin{array}{l}
\alpha_1 f(y_i;\bm{X}_{1i},\bm{X}_{2i},\bbeta_k,\bm{\eta}), \qquad \text { at existing } k \\
\alpha_2g(y_i;\tau_0,\tau,\Sigma_0,\Sigma,a_0,b_0,\bm{X}_{2i},\bm{\eta}), \qquad \text {if } k \text { is a new cluster }
\end{array}\right.
\label{eq:Cluster}
\end{align*}
where
\begin{align*}
   &f(y_i;\bm{X}_{1i},\bm{X}_{2i},\bbeta_k,\bm{\eta}) = \frac{1}{(2 \pi \sigma_k^2)^{1/2}} \exp\big( -\frac{1}{2 \sigma_k^2} \|y_{i}-\bm{X}_{1i} \bbeta_k-\bm{X}_{2i} \boldsymbol{\eta}\|^2 \big\}\\
   &g(y_i;\tau_0,\tau,\Sigma_0,\Sigma,a_0,b_0,X_{2i},\bm{\eta})=
   \frac{b_0^{a_0} \Gamma(a_0 + \frac{1}{2}) |\bSigma|^{1/2}}{(2 \pi)^{1/2} \Gamma(a_0) |\bSigma_0|^{1/2}} \\
   &\times \big\{ b_0+\frac{1}{2} ( \btau_0^\top \bSigma_0^{-1} \btau_0 + (y_{i}-\bm{X}_{2i} \boldsymbol{\eta})^2 - \btau\bSigma^{-1}\btau ) \big\} ^{-(a_0 + \frac{1}{2})}\\
   &{\btau} = \bSigma(\bSigma_0^{-1}\btau_0 + \bm{X}_i (y_{i}-\bm{X}_{2i} \boldsymbol{\eta})),\\
   &{\bSigma} = (\bSigma_0^{-1} + \bm{X}_i^\top \bm{X}_i)^{-1}.
\end{align*}
\end{prop}
Proposition \ref{prop1} shows that we can directly sample $z_1,\ldots,z_n$ interactively from a categorical distribution.


\begin{prop}
\label{prop2}

The full conditional distribution of $(\boldsymbol{\beta}_{k},{\sigma}_k^2)$ is given as 
\begin{align*}
    P(\boldsymbol{\beta}_{k},{\sigma}_k^2|\bm{z},\bm{\eta})  & \propto \Big(\frac{1}{\sigma_k^2}\Big)^{a_0+\frac{N_c}{2} + \frac{1}{2}+1} \exp\Big[ - \frac{1}{\sigma_k^2} \Big\{ b^* + \frac{1}{2} (\bbeta_k - \btau^*)^\top \bSigma^{* {-1}} (\bbeta_k - \btau^*) \Big\} \Big] \\
    & \propto \text{NIG}(\btau^*, \bSigma^*, a^*, b^*),
\end{align*}
where 
\begin{align*}
    & \btau^* = (\bSigma_0^{-1} + \sum_{z_i = k} \bm{X}_{1i}^\top \bm{X}_{1i})^{-1} (\bSigma_0^{-1}\btau_0 + \sum_{z_i = c} \bm{X}_{1i} (y_i-\bm{X}_{2i}\bm{\eta})), \\
    & \bSigma^* = (\bSigma_0^{-1} + \sum_{z_i = c} \bm{X}_{1i}^\top \bm{X}_{1i})^{-1} ,\\
    & a^* = a_0 + N_c/2, 
    \\
    &b^* = b_0 + \frac{1}{2}[\btau_0^\top \bSigma_0^{-1} \btau_0 + \sum_{z_i = c} (y_{i}-\bX_{2i}\boldsymbol{\eta})^2 - \btau^{*\top} \bSigma^{* {-1}} \btau^*],\\
    &\text{and} \ N_c = \sum_i I(z_i = c).
\end{align*}
\end{prop}  

Proposition \ref{prop2} shows that given the cluster memberships $\bm{z}$ and $\bm{\eta}$, $(\boldsymbol{\beta}_{k},{\sigma}_k^2)$ can be sampled from a normal-inverse-gamma distribution. The posterior distribution of $\tilde \beta$'s is obtained by $\tilde \beta_k= \tilde{\bm{M}}_1\bm{\beta}_k$.

\begin{prop} \label{prop3}  The full conditional distribution of $\bm{\eta}$ is given as 
\begin{align*}
    P(\bm{\eta}|\bm{z},\bm{\beta},\bm{\sigma}^2) & \propto |\bm{V}^{*}|^{-1/2} \exp \Big\{  \frac{1}{2} (\bm{\eta} - \bm{\eta}^*)^\top \bm{V}^{* {-1}} (\bm{\eta} - \bm{\eta}^*) \Big\} \\
    & \propto \mathcal{N}_p(\bm{\eta}^*, \bm{V}^{*}),
\end{align*}
where 
\begin{align*}
& \bm{\eta}^* = (\bm{V}_0 + \sum_{i = 1}^n \bm{X}_{2i} \bm{X}_{2i}^\top)^{-1} (\bm{V}_0\bm{\eta}_0 + \sum_{i = 1:n} \bm{X}_{2i}^\top (y_i-\bm{X}_{1i}^\top\bm{\beta_{z_i}})), \\
    & \bm{V}^* = (\bm{V}_0 + \sum_{i = 1}^n \bm{X}_{2i} \bm{X}_{2i}^\top)^{-1}. 
\end{align*}
\end{prop}
The proof of the three Propositions are given in the supplementary materials. 
We summarize the collapsed Gibbs sampling procedure for the proposed model in Algorithm \ref{algo:collapsed}.


\begin{algorithm}
\caption{Collapsed Gibbs sampler}
\label{algo:collapsed}
\begin{algorithmic}[1]
\State  Observations $y_{1},...,y_{n}$
\State Initialize the cluster number $K$, the cluster configuration $\bm{z}$, and the parameters $\bm{\beta},\bm{\sigma}^2,\bm{\eta}$ for each $y_{i}$
\For { $iter$ from $1$ to $M$ }
       \For{$i$ from $1$ to $N$ }
       \State Update $\bm{z}$ conditional on $\bm{\beta},\bm{\eta},\bm{\sigma}^2$ for each $i$ by Proposition \ref{prop1}
       \EndFor
       \For{$c$ from 1 to $K$} 
       \State Update $a,b,\sigma^2_c,\bm{\beta}_c$ for each $c$ conditional on $z$ by Proposition \ref{prop2}.
       \EndFor
       \State Update $\bm{\eta}$ conditional on $\bm{z},\bm{\beta},\bm{\sigma}^2$ by Proposition \ref{prop3}.
\EndFor
\end{algorithmic} 
\end{algorithm}

\subsection{Post MCMC Estimation}
The posterior inference on the cluster memberships $z_1,\ldots,z_n$ is carried out by Dahl's method \citep{dahl2006model}. We first need to define the membership matrix which contains all the information of cluster configurations. Let $\bm{z}^{(m)}=\{z_{i}^{(m)}, 1\leq i \leq n\}$ be the $m$ th estimation of cluster configurations after the burn-in iterations. The membership matrix for the $m$ th posterior sample is defined as
\begin{align*}
\bm{B}^{(m)}=[b^{(m)}_{ij}]_{n\times n}=[I(z^{(m)}_{i}=z^{(m)}_{j})]_{n\times n}, m=1,...,M ,
\end{align*}
where $I$ is the indicator function.  Thus, if state $i$ and state $j$ are in the same cluster in the $m$ th posterior sample, $b_{ij}^{(m)}=1$. If not, $b_{ij}^{(m)}=0$. Then, we can define the mean of all  membership matrices after burn-in iterations as
\begin{align*}
    \bar{\bm{B}}=\frac{1}{M}\sum_{m=1}^{n}\bm{B}^{(m)}.
\end{align*}
The best posterior sample is selected by the least square distance to $\bar{\bm{B}}$:
\begin{align*}
   m_{best}={\text{argmin}}_{m\in \{1:M\}}\left\|\bm{B}^{(m)}-\bar{\bm{B}}\right\|,
\end{align*}
where $||\bm{B}^{(m)}-\bar{\bm{B}}||=\sum_{i=1}^{n}\sum_{j=1}^{n}(\bm{B}^{(m)}(i,j)-\bar{\bm{B}}(i,j))^2$. Therefore, the posterior estimates of cluster memberships
$z_1,\ldots,z_n$ and cluster-wise parameters can be obtained
based on the draw identified by Dahl’s method.


\subsection{Model Selection Criterion}

We recast the choice of decaying parameter $\lambda $as a model
selection problem. We use the Logarithm of the Pseudo-
Marginal Likelihood \citep[LPML;][]{ibrahim2013bayesian} based on
conditional predictive ordinate  \citep[CPO;][]{gelfand1994bayesian} to select $\lambda$. The LPML given $\lambda$ is defined as
\begin{gather}
	\text{LPML}=\sum_{i=1}^{n}\text{log}\{\text{CPO}_i(\lambda)\},
\end{gather}
where $\text{CPO}_i(\lambda)$ is the ith CPO under a certain $\lambda$ value.
Following \citet{chen2012monte}, a Monte Carlo estimate of the CPO can be obtained as 
\begin{align*}
     \widehat{\text{CPO}}_i(\lambda)=\{\frac{1}{M}\sum_{m=1}^M\frac{1}{L(\bm{\theta}^{(m)}_{z_i};\lambda)}\},
\end{align*}
where $M$ is the total number of Monte Carlo iterations after burn-in, $\bm{\theta}^{(m)}_{z_i}$ is the mth posterior sample and $L(\cdot;\lambda)$ is the likelihood function. Then, an estimate of the LPML can subsequently be
calculated as
\begin{align*}
    \widehat{\text{LPML}}(\lambda)=\sum_{i=1}^{n}\text{log}\{ \widehat{\text{CPO}}_i(\lambda)\}.
\end{align*}
A model with a larger LPML value indicates a preferred model.

\section{Simulation}\label{sec:simu}
In this section, we detail the simulation settings, the evaluation metrics, and the comparison performance results.
\subsection{Simulation Settings and Performance Measurements}
\begin{figure}[H] 
	\centering  
	\subfigure[]{
		\includegraphics[scale=0.3]{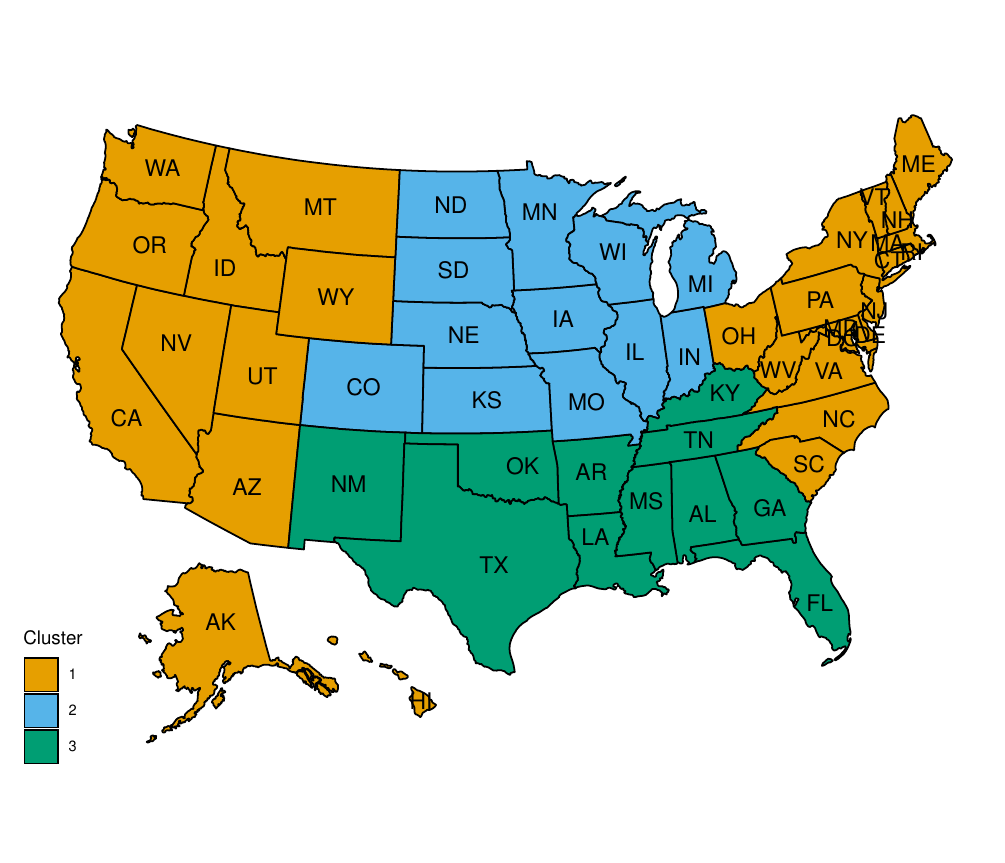}}
	\subfigure[]{
		\includegraphics[scale=0.3]{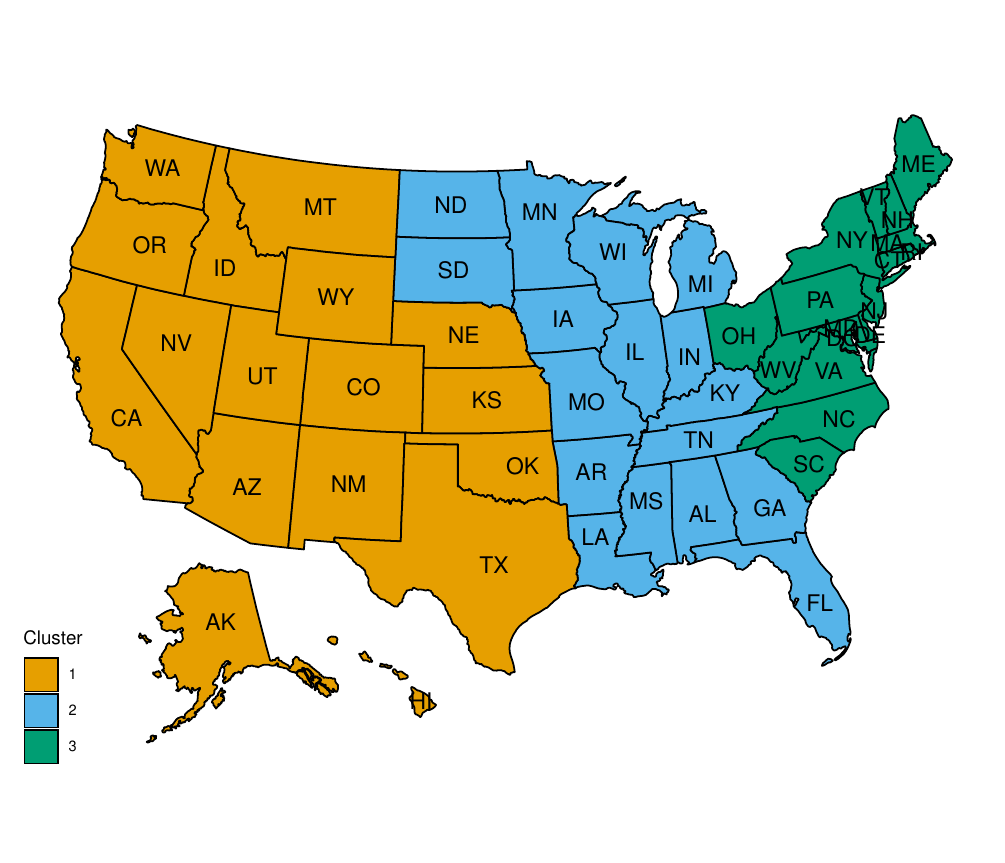}}
	\caption{Illustration of two partitions for simulation
	\label{fig:partition} }
\end{figure}

The spatial adjacency structure of the 51 states is used in the data simulation. We
consider in total two partition settings which are shown in Figure \ref{fig:partition}.  The true number of clusters for both settings is 3. The first partition setting shown in
Figure~\ref{fig:partition} (a) contains a cluster which consists of two
disjoint parts in the east and west. It is designed to mimic a fairly common
economic pattern. There are no disjoint components in second partition design, which is shown in Figure~\ref{fig:partition} (b).

\begin{table}[H]
\centering
\caption{Two parameter settings in the simulation study}\label{tab:simu_set}
\footnotesize
\resizebox{\columnwidth}{1in}{
\begin{tabular}{c|c|c|c|c|c}
    \toprule Cluster&$\tilde{\bbeta}$&$\bm{X}_{1i}$&$\bm{\eta}$&$\bm{X}_{2i}$&$\epsilon_{i}$  \\
    \midrule
    \multicolumn{6}{c}{Parameter Setting 1}\\
    \midrule
    1&(1, -2, 1)&\multirow{3}{*}{Dir (1, 3, 6)}&\multirow{3}{*}{(1, 2, 1)}&\multirow{3}{*}{$U(-1,1)$}&\multirow{3}{*}{$N(0,1)$}\\
    \cline{1-2}2&(-4, -3, 7)&&&&\\
    \cline{1-2}3&(10, -9, -1)&&&&\\
    \midrule
    \multicolumn{6}{c}{Parameter Setting 2}\\
    \midrule
    1&(1, 1, 1, 1, 1, -1, -1, -1, -1, -1)&\multirow{3}{*}{Dir (1, 4, 5, 3, 8, 7, 1, 3, 2, 6)}&\multirow{3}{*}{(1, 2, 1)}&\multirow{3}{*}{$U(-10,10)$}&\multirow{3}{*}{$N(0,1)$}\\
    \cline{1-2}2&(-2, 5, -3, -2, 5, -3, -3, 6, -1, -2)&&&&\\
    \cline{1-2}3&(3,  -3, -2, 8, -4, -2, 8, -2, -4, -2)&&&&\\
    \bottomrule
\end{tabular}
}

\end{table}
For data generation, we consider the following data generation model:
\begin{gather}
    y_{i}=\bm{X}_{1i}\tilde{\bbeta}_{z_{i}}+\bm{X}_{2i}\bm{\eta}+\epsilon_{i},\qquad  \text{s.t.}\quad \sum_{j=1}^k\tilde{\beta}_{z_i,j}=0
\end{gather}
where $\tilde{\bbeta}_{z_i}=[\tilde{\beta}_{z_i,1},...,\tilde{\beta}_{z_i,k}]$. The true values
of the parameters are set as Table \ref{tab:simu_set}. We consider two different parameter settings with different dimensions of compositional covariates. In the first simulation setting, the compostional data is generated from a 3-dimensional Dirichlet distribution with parameters $(1, 3, 6)$. 
A total of 100 datasets are generated for each simulation setting. For each replicate, the proposed approach is applied with different $\lambda$ values ranging from 0 to 5 with a step size of 0.5, and the best $\lambda$ is chosen through LPML. A total of 1500 MCMC iterations are run for each replicate, with the first 500 iterations are set as burn-in.

The estimated number of clusters $\hat{K}$ and rand index (RI) \citep{rand1971objective} are calculated to evaluate the clustering performance. The RI is calculated using the final clustering result obtained by Dahl's method for each replicate, and we report the average RI over all replicates.  Let $\beta_{lm}$ denote the true value of  $m$ th coefficient of $\tilde{\bbeta}$ in state $l$, and $\hat{\beta}_{lmr}$ denote the posterior estimate for $\beta_{lm}$ in the $r$th replicate. Then the estimation performance of $\tilde{\bbeta}$ will be evaluated by mean absolute bias (MAB), mean standard deviation (MSD), and mean of mean squared error (MMSE) as
\begin{equation*}
\begin{split}
      \text{MAB}=\frac{1}{51}\sum_{l=1}^{51}\frac{1}{100}\sum_{r=1}^{100}|\hat{\beta}_{lmr}-\beta_{lm}|,\\
    \text{MSD}=\frac{1}{51}\sum_{l=1}^{51}\sqrt{\frac{1}{99}\sum_{r=1}^{100}(\hat{\beta}_{lmr}-\bar{\hat{\beta}}_{lm})^2},\\
    \text{MMSE}=\frac{1}{51}\sum_{l=1}^{51}\frac{1}{100}\sum_{r=1}^{100}(\hat{\beta}_{lmr}-\beta_{lm})^2. 
\end{split}
\end{equation*}

Similarly, the MAB, MSD, MMSE for $\bm{\eta}$ is defined as
\begin{equation*}
\begin{split}
    \text{MAB}=\frac{1}{100}\sum_{r=1}^{100}|\hat{\eta}_{mr}-\eta_{m}|,\\
    \text{MSD}=\sqrt{\frac{1}{99}\sum_{r=1}^{100}(\hat{\eta}_{mr}-\bar{\hat{\eta}}_{m})^2},\\
    \text{MMSE}=\frac{1}{100}\sum_{r=1}^{100}(\hat{\eta}_{mr}-\eta_{m})^2
\end{split}
\end{equation*}
where $\eta_{m}$ is the true value of  $m$th coefficient of $\bm{\eta}$, $\hat{\eta}_{mr}$ denote the posterior estimate for $\eta_{m}$ in the $r$th replicate.

\subsection{Simulation Result}

\begin{figure}[H]
\centering
\subfigure{\includegraphics[width=3.0in]{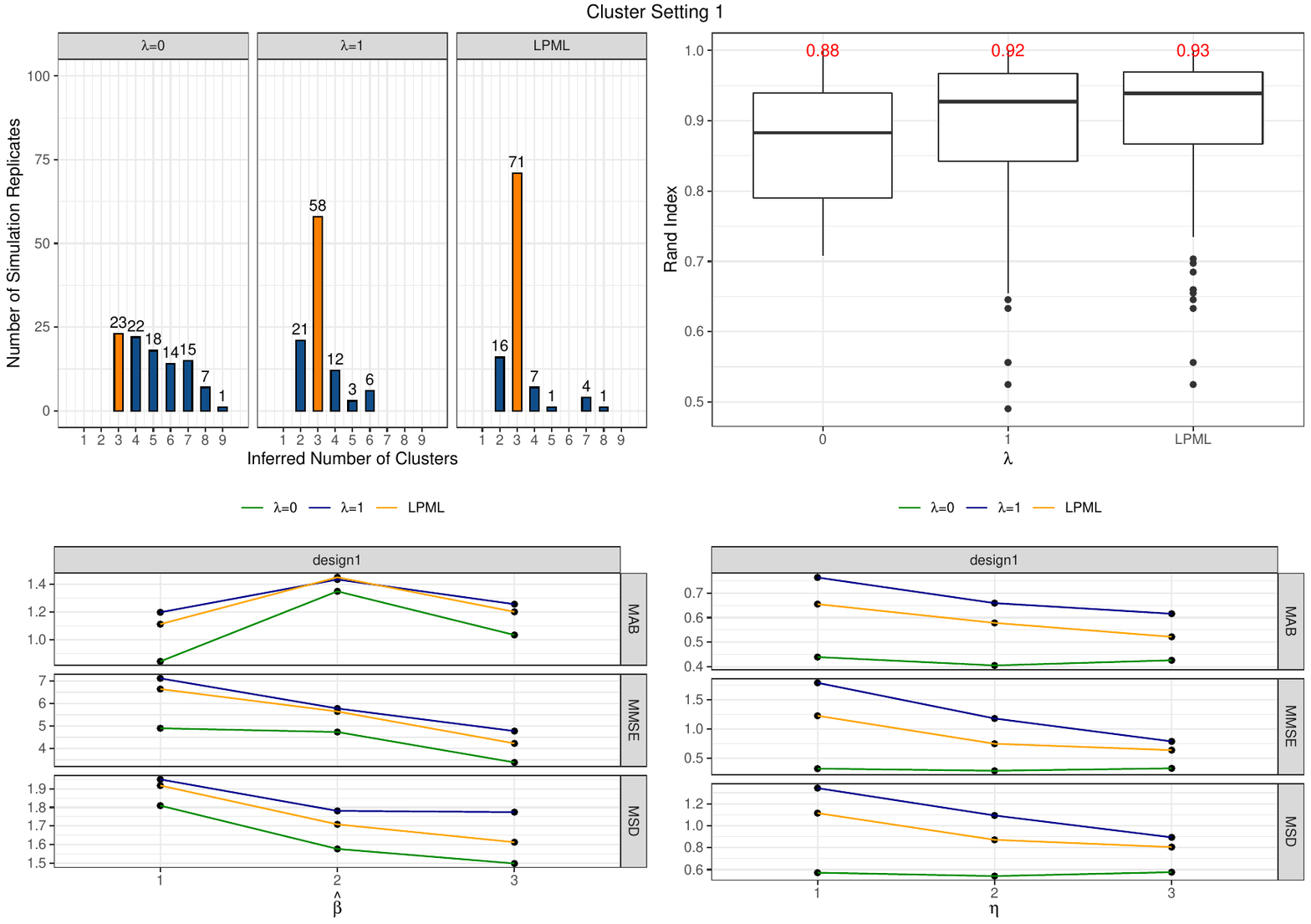}}
\subfigure{\includegraphics[width=3.0in]{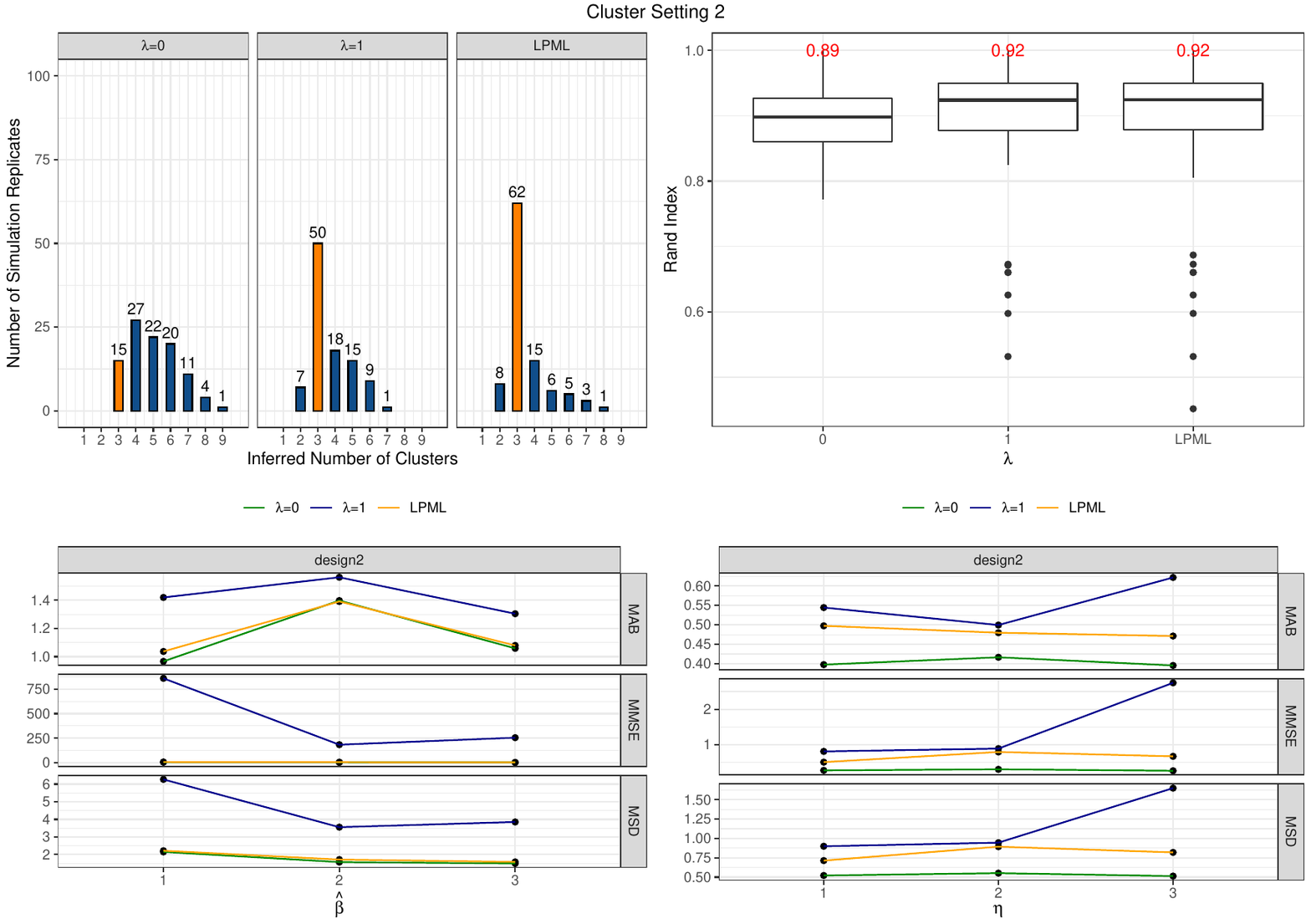}}
\caption{Results of performance evaluation under parameter setting 1. Left top panel: histogram of clusters. Right top panel: box plot of rand index.  Left bottom panel: MAB, MSD, MMSE of $\hat{\beta}$.  Right bottom panel:  MAB, MSD, MMSE of $\eta$. The red texts show the median of rand index.} 
\label{fig:Parameter setting 1}
\end{figure}

\begin{figure}[H]
\centering
\subfigure{\includegraphics[width=3.0in]{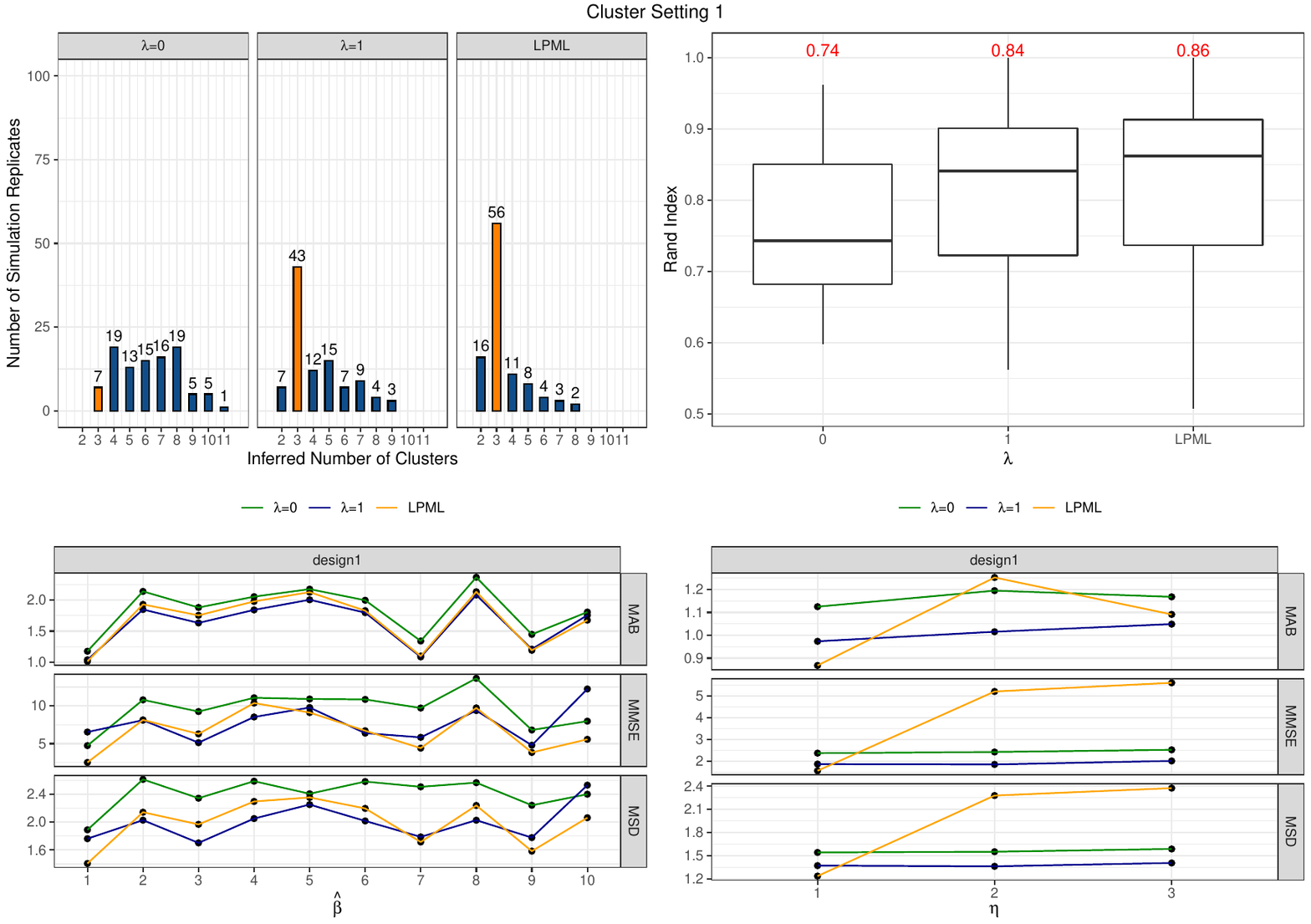}}
\subfigure{\includegraphics[width=3.0in]{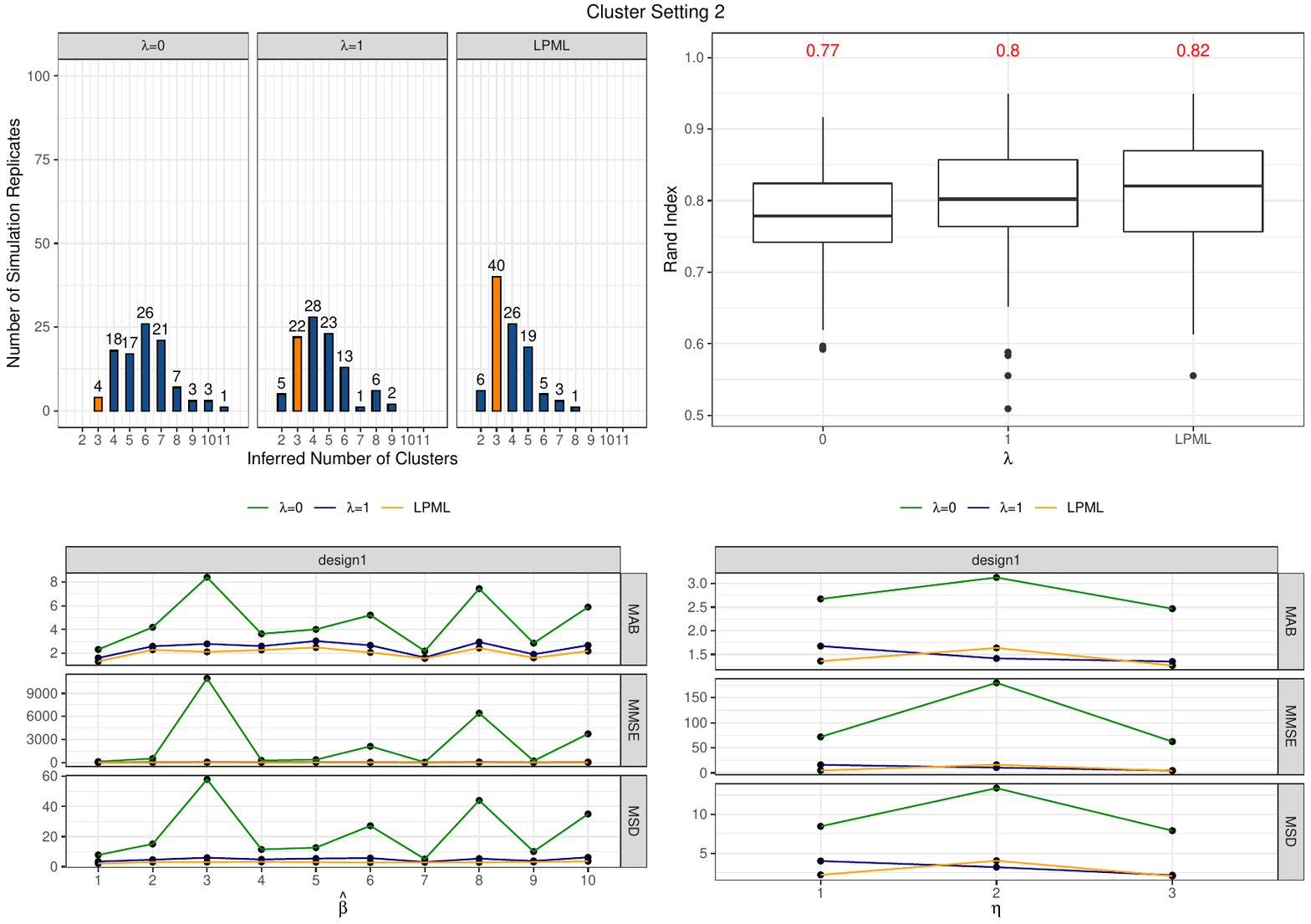}}
\caption{Results of performance evaluation under parameter setting 2. Left top panel: histogram of clusters. Right top panel: box plot of rand index.  Left bottom panel: MAB, MSD, MMSE of $\hat{\beta}$.  Right bottom panel:  MAB, MSD, MMSE of $\eta$. The red texts show the median of rand index.} 
\label{fig:Parameter setting 2}
\end{figure}

We consider $\lambda\in \{0, 0.5, 1.0, 1.5, 2.0, 2.5, 3.0, 3.5, 4.0, 4.5, 5.0\}$, and the best $\lambda$ value is selected by LPML.
Clustering and estimating performances are visualized in Figure~\ref{fig:Parameter setting 1} and Figure~\ref{fig:Parameter setting 2}. When the best $\lambda$ is selected by LPML, the medians of rand index of first and second partitions are 0.93 and 0.92, respectively. For both partitions, the number of clusters inferred correctly and the average rand index are better than MFM. We see that MAB, MMSE, MSD for $\tilde{\bbeta}$ and $\bm{\eta}$ are small for both design 1 and design 2, which indicates a good fit of $\tilde{\bbeta}$ and $\bm{\eta}$. And for parameter setting 1, the MAB, MMSE, MSD for $\tilde{\bbeta}$ and $\bm{\eta}$ of $\lambda$ selected by LPML is larger than that of $\lambda = 0$. When $\lambda$ is selected by LPML, we will have less small clusters and have more replicates with cluster number less than true cluster number.  And when cluster number inferred is less than true cluster number, the MAB, MMSE, MSD will be higher. This might be the partial reason why the MAB, MMSE, MSD of $\lambda$ selected by LPML is higher than $\lambda = 0$.


From Figure \ref{fig:Parameter setting 2}, we see that the medians of rand index of first and second partitions are 0.86 and 0.82, respectively. When $\lambda=0$, we can see that the correct inferred number of clusters is very low although the rand index is acceptable. Under our proposed model, both the correct inferred number of clusters and the rand index increase a lot. Different from parameter setting 1, the performance of estimation for $\tilde{\bbeta}$ with $\lambda$ selected from LPML is better than that with $\lambda = 0$. The partial reason might be that the corrected number of clusters is too low when $\lambda = 0$, which leads to high MAB, MSD, and MMSE of both $\tilde{\bbeta}$  and $\bm{\eta}$.

In a brief conclusion based on our simulation studies, the proposed models have better performance than MFM both for clustering and parameter estimation. Our proposed model
selection criterion, LPML, can nearly select the best performing $\lambda$ value for both clustering
and parameter estimation.

\section{Linking intersectoral GDP contributions to Gini Coefficients}\label{sec:app}

In this section, the proposed approach is applied to explore the spatial homogeneity among states according to the regression of the intersectoral GDP contributions in the U.S. to the Gini coefficient. We take other demographics information of 51 states into consideration,  including household income per capital, unemployment rate.

\begin{figure} [t]
	\centering
	\includegraphics[scale=0.45]{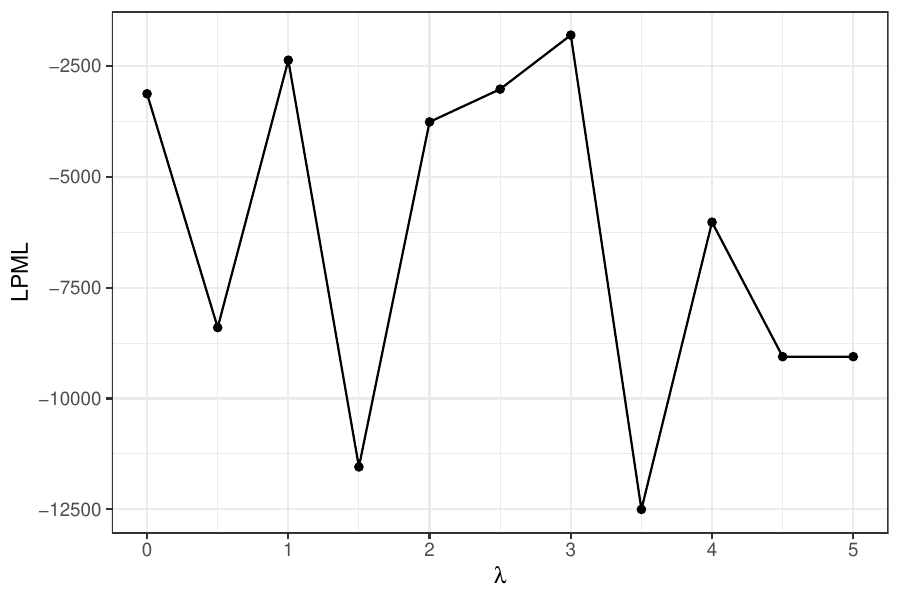}
	\caption{LPML values of different $\lambda$ values} \label{fig:LPML}
\end{figure}

First, we construct the adjacency matrix among the states. Specifically, the adjacency matrix is defined by the binary matrix $\bm{W}$ where $W_{ij}=1$ if state $i$ and state $j$ share some common boundary and $W_{ij}=0$ otherwise. The graph distance \citep[GD;][]{bhattacharyya2014community} is used
as the distance measure to construct the neighborhood graph used in the Markov
random field model. The upper limit of the graph distance for two states is set to~2 to be
considered as ``neighbors''.
The spatial smoothing parameter~$\lambda$ is considered within
the range of \{0, 0.5, 1, \ldots, 5\}, with $\lambda = 0$ corresponding to
MFM model. All hyperparameters are set to be the same as in simulation studies. To implement our method, we run MCMC chain with $1000$ total iterations where the first $500$ draws are discarded as burn-in to ensure the convergence. The LPML is used to determine the
optimal spatial smoothing parameter.
Figure \ref{fig:LPML} shows the LPML values with different values of $\lambda$. It is shown that when $\lambda = 3.0$, the LPML value turns out to be the largest, indicating that the model selection criterion chooses $\lambda = 3.0$ as the best value. 

\begin{figure} [H]
\centering
	\subfigure[]{
	\includegraphics[scale=0.44]{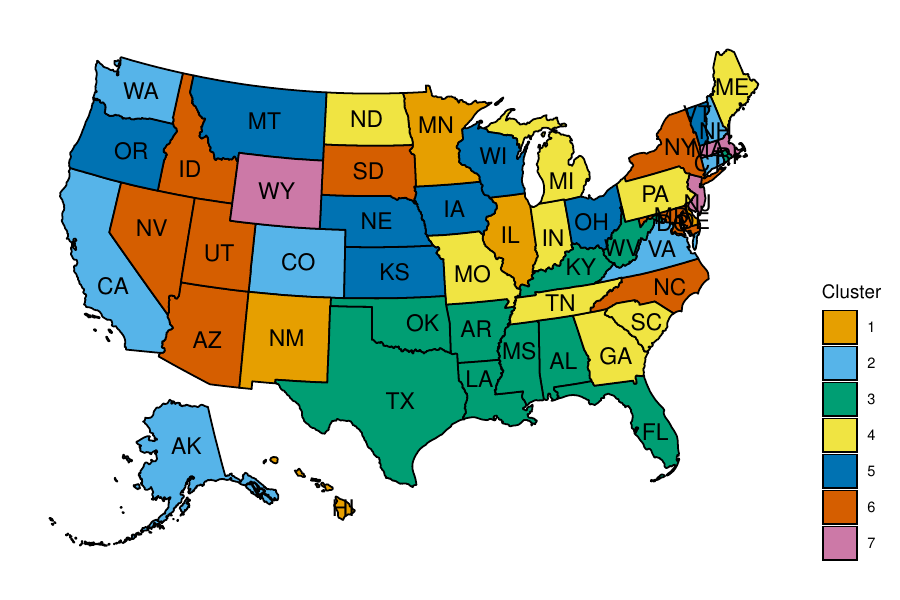}
	}
	\subfigure[]{
	\includegraphics[scale=0.44]{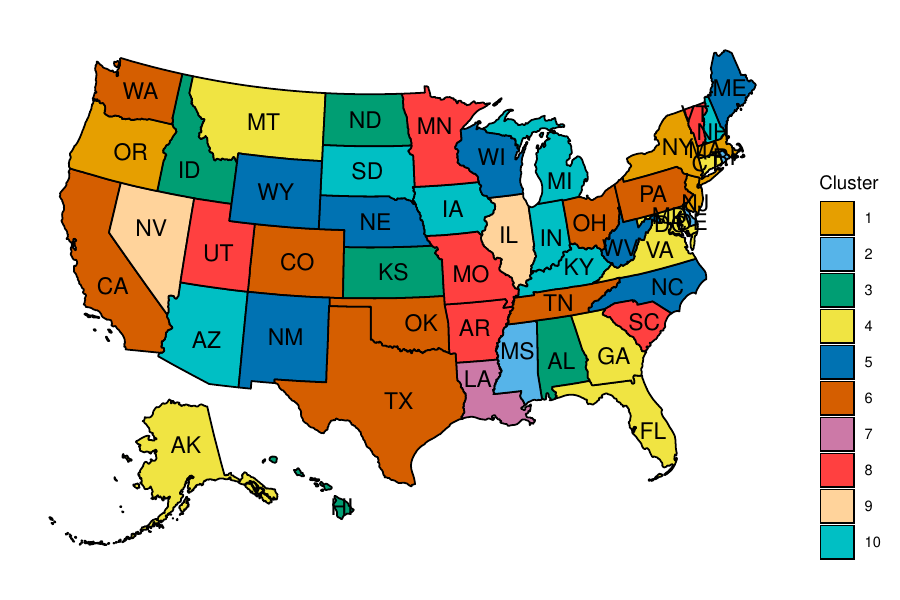}
	}
	\caption{(a) Clustering results using the proposed method for the 51 states with $\lambda$ selected from LPML. (b) Clustering results for the 51 states with $\lambda = 0$. } \label{fig:realcluster}
\end{figure}

 After applying Dahl's method, we obtain 7 clusters of size 5, 7, 10, 10, 8, 8 and 3, respectively. The final cluster configuration is visualized in
Figure~\ref{fig:realcluster}. Cluster~3 is the highest in terms of income inequality,
and has an average Gini coefficient of 0.47701. Cluster~5, with an average Gini of
0.4491, exhibits the most equal income distribution among the seven.  We compare the clustering result by setting $\lambda = 0$, whose estimation result is shown in Figure \ref{fig:realcluster} (b). We observe that there are less small clusters when using $\lambda$ selected from LPML compared to  $\lambda =0 $. A significant advantage of our proposed method is that it allows for globally
discontinuous clusters. As illustrated in Figure~\ref{fig:realcluster}, most southern states are clustered into one cluster. Their Gini coefficient are relatively high. Kentucky and West Virginia are also included in this cluster. Both have high Gini Coefficient, low income per capital and high unemployment rate. Most states of Cluster 4 are located at Midwest and South. These states also have relatively high Gini coefficient.

The estimated coefficients for each cluster is shown in Table \ref{table:estimation for coefficients}. From Table~\ref{table:estimation for coefficients}, we see that $\hat{\tilde{\beta}}_{4}$
 is negative for all clusters except Cluster 3, which indicates that the increasing of construction will lead to lower income inequality for most states. In addition, $\hat{\tilde{\beta}}_{8}$ is negative for states in Cluster~2,~3,~4 and,~7 and $\hat{\tilde{\beta}}_{8}$ is also negative for Cluster~2,~3,~6,~7. Those indicate that the increasing of Transportation and warehousing will improve the income inequality for states with worse income inequality. For those states, expanding Educational services, health care, and social assistance will improve the equality of income distribution. Furthermore, $\hat{\tilde{\beta}}_{12}$ is negative for states in Cluster~2, 3, and 5 which have high average Gini coefficient. Increasing investment on educational services, health care, and social assistance is a strategy for improving the equality of income distribution. 


\begin{sidewaystable}
\scriptsize
\caption{Estimation for parameters of each cluster}
\begin{tabular}{ccccccccc}
\toprule
 && Cluster 1 &Cluster 2 &Cluster 3
&Cluster 4
&Cluster 5
&Cluster 6
&Cluster 7
\\
\midrule
  $\hat{\tilde{\beta}}_1$&Agriculture & -0.045& 0.07& -0.168& -0.277& -0.068& -0.117& 0.003
\\
$\hat{\tilde{\beta}}_2 $&Mining
& -0.077& -0.104& -0.052& 0.249& -0.096& -0.102& 0.032\\
 $\hat{\tilde{\beta}}_3$& Utilities
& -0.429& -0.054& -0.214& 0.115& 0.048& 0.442& 0.395
\\
 $\hat{\tilde{\beta}}_4$&Construction & -0.078& -0.132& 0.142& -0.125& -0.147& -0.337& -0.216\\
 \midrule
$ \hat{\tilde{\beta}}_5$&Manufacturing & 0.054& 0.44& -0.276& -0.135& -0.117& -0.339& 0.044\\

 $\hat{\tilde{\beta}}_6$&Wholesale 
& 0.061& 0.098& -0.070& -0.171& 0.262& -0.401& 0.055\\
 $\hat{\tilde{\beta}}_7$&Retail & 0.307& 0.006& 0.296& 0.062& 0.02& -0.066& -0.668\\

 $\hat{\tilde{\beta}}_8$&Transportation & 0.137& -0.301& -0.001& -0.231& 0.207& 0.079& -0.567\\
\midrule
 $\hat{\tilde{\beta}}_9$&Information & -0.02& 0.28& -0.122& 0.003& -0.186& 0.413& 0.801\\

 $\hat{\tilde{\beta}}_{10}$&Finance & -0.117& 0.289& -0.07& 0.038& 0.117& 0.253& 0.469\\

 $\hat{\tilde{\beta}}_{11}$&Profession service & -0.064& 0.285& -0.073& 0.261& -0.035& 0.097& -0.053\\

 $\hat{\tilde{\beta}}_{12}$&Education, Health Care & 0.279& -0.29& -0.101& 0.221& 0.137& -0.32& -0.372\\
\midrule
 $\hat{\tilde{\beta}}_{13}$&Entertainment & 0.044& -0.175& -0.06& 0.03& -0.008& -0.16& -0.476\\
 $\hat{\tilde{\beta}}_{14}$&Other service & -0.236& -0.355& 0.368& 0.372& 0.049& -0.03& -0.293\\

 $\hat{\tilde{\beta}}_{15}$&Federal Spending & 0.176& -0.16& 0.071& -0.459& -0.052& 0.236& 0.313\\

 $\hat{\tilde{\beta}}_{16}$&Local Spending& 0.009& 0.103& 0.328& 0.047& -0.133& 0.354& 0.532\\
\midrule
 $ \hat{\eta}_1$&Household Income &\multicolumn{7}{c}{
  [-0.562, 0.523]}
\\

$\hat{\eta}_2$& Unemployment Rate&\multicolumn{7}{c}{[-0.413, 0.532]}
\\
\midrule
$\hat{\sigma}^2$
&[ 0.066 ,2.375 ]
&[ 0.071 ,11.654 ]
&[ 0.07 ,24.822 ]
&[ 0.071 ,2.288 ]
&[ 0.061 ,2.597 ]
&[ 0.069 ,1.175 ]
&[ 0.078 ,7.395 ]
\\
\bottomrule
\end{tabular}

\label{table:estimation for coefficients}
\end{sidewaystable}


Finally, to make sure the cluster configuration presented here is not a random
occurrence but reflects the true pattern demonstrated by the data, we ran 100
separate MCMC chains with different random seeds and initial values, and
obtained 100 final clustering schemes. The RI between each scheme and the
present clustering scheme in Figure~\ref{fig:realcluster} is calculated, and they
average to 0.78.

\section{Discussion}\label{sec:discussion}
In this study, we introduced a novel Bayesian log-contrast compositional regression approach for deciphering the heterogeneity pattern inherent in compositional data. This innovative method allows the estimation of the unknown parameters within the regression for compositional covariates, and also facilitates the inference of both the cluster configuration and the number of clusters present. To accomplish Bayesian inference, we developed an efficient Markov Chain Monte Carlo (MCMC) algorithm. We further utilized Dahl's method to carry out the post-MCMC estimation and adopted the logarithm of the Pseudo-Marginal Likelihood based on the conditional predictive ordinate for tuning parameter selection. In comparison to conventional techniques such as Support Vector Clustering Machines (SVCMs), our proposed method eliminates the need for pre-specifying the number of clusters and effectively utilizes spatial information. It ensures the maintenance of local contiguous constraints and global discontinuous clusters, provides easy interpretations of clustering results, and facilitates efficient posterior inference of cluster-wise parameters and clustering information. The numerical results showed that the proposed method can accurately estimate cluster information and cluster-wise parameters under different scenarios.
 A case study using the
BEA data reveals a number of important findings for relationships between Gini coefficients and GDP contributions across the~51 states in
the US. The results provide valuable insights to both
residents and governors: residents could gain a better understanding of the economic development of their
state's conditions, and hence vote with their feet accordingly; governors,
equipped with more \emph{objective} and principled analysis of income inequality, could make
better data-informed policy design decisions to switch their industrial structure.

A few topics beyond the scope of this paper are worth further investigation. Our methodology involving the log-ratio transformation of compositional data necessitates the elements of the covariance matrix to be nonzero. We employ a small number as a substitute in instances where zero elements are present before executing the log-ratio transformation. However, this approach may prove inadequate in a high-dimensional model context, where the prevalence of zero elements is significantly higher, and the remaining nonzero elements are potentially minimal. Consequently, substituting zero elements with minuscule numbers may be ineffective within such a model configuration. Therefore, the implementation of more advanced high-dimensional techniques may be required to address this challenge adequately. In addition, nonstationarity is an essential
consideration of spatial dependence. Taking a nonstationary cost function into account can broaden the applications in various domains. 

\bibliographystyle{chicago}
\bibliography{mfm-com.bib}

\end{document}